\documentclass[pra,onecolumn,tightenlines,10pt,amsfonts,amssymb,balancelastpage,notitlepage]{revtex4-1}%
\usepackage[utf8x]{inputenc}

\usepackage{graphics}
\usepackage{epsfig}
\usepackage{subfigure}
\usepackage{rotating}
\usepackage{mathrsfs}
\usepackage{amsfonts}
\usepackage{amsmath}
\usepackage{amssymb}
\usepackage{placeins}
\usepackage{bm}

\usepackage{pgfplots}
\usepackage{tikz}
\usetikzlibrary{arrows,decorations.markings}
\pgfplotsset{compat=1.14}

\usepackage{soul}
\usepackage{fixmath}
\usepackage{physics}

\usepackage[colorlinks]{hyperref}

\newcommand{\hc}{\ensuremath{\text{h.c.}}}
\newcommand{\transp}{\ensuremath{\mathrm{T}}}

\newcommand{\latin}[1]{{\emph{#1}}}
\DeclareMathOperator{\diag}{diag}

\begin{document}

\title{Generalized approach for enabling multimode quantum optics}
\author{Élie Gouzien$^{1}$, Sébastien Tanzilli$^{1}$, Virginia D'Auria$^{1}$, and Giuseppe Patera$^{2}$}

\address{
$^1$Universit\'{e} Cote d’Azur, CNRS, Institut de Physique de Nice, Parc Valrose, 06108 Nice Cedex 2, France\\
$^2$Univ. Lille, CNRS, UMR 8523 - PhLAM - Physique des Lasers Atomes et Mol\'{e}cules, F-59000 Lille, France
}

\begin{abstract}
We develop a universal approach enabling the study of any multimode quantum optical system evolving under a quadratic Hamiltonian.
Our strategy generalizes the standard symplectic analysis and permits the treatment of multimode systems even in situations where 
traditional theoretical methods cannot be applied. This enables the description and investigation of a broad variety of key-resources 
for experimental quantum optics, ranging from optical parametric oscillators, to silicon-based micro-ring resonator, as well as opto-mechanical systems.
\end{abstract}
\pacs{vvv}

\date{\today}
\maketitle
\section{Introduction}
Multimode quantum optics in continuous variables (CV) regime is at the hearth of a many  quantum applications, 
encompassing quantum communication~\cite{Weedbrook2012,Ferraro_book}, quantum metrology~\cite{Giovannetti2004} as well as quantum computation~\cite{Braunstein2005a} 
via cluster states~\cite{Zhang2006,Menicucci2006, Furusawa2016}.

A central step in the treatment of multimode optical systems lies in the identification of the so-called super-modes~\cite{Wasilewsky2006, Valcarcel2006, Patera2010}. 
These are coherent superpositions of the original modes, that diagonalize the equations describing the system dynamics, 
and permit to rewrite multimode CV entangled states as a collection of independent squeezed states~\cite{Braunstein2005}.
The knowledge of super-modes is required to optimise the detection of the non-classical information on the state~\cite{Bennink2002,Valcarcel2006,Wasilewsky2006}, 
to generate and exploit CV cluster states in optical frequency combs~\cite{Menicucci2007,Patera2012,Roslund2014} or in multimode spatial systems~\cite{DavidBarral2019} 
as well as to engineer complex multimode quantum states~\cite{Ferrini2015,Nokkala2018}. In experiments, as they are statistically independent, 
super-modes can be measured with a single homodyne detector, thus considerably reducing the experimental overhead~\cite{Roslund2014}.

Due to their multiple purposes, a universal strategy allowing to retrieve the super-modes is crucial for multimode quantum optics and its applications. 
The aim of this theoretical work is to provide such a powerful and versatile tool. 
More specifically, multimode optical quantum states are generally produced via a non-linear interaction described by a quadratic Hamiltonian~\cite{Ferraro_book}.
The transformation diagonalizing the system equations must be symplectic, \latin{i.e.\@} preserve the commutation rules. Standard methods for symplectic diagonalization, 
such as Block-Messiah decomposition (BMD)~\cite{Arvind1995}, are valid for single pass interactions~\cite{Adesso2014,Christ2011,Lipfert2018} 
but not suitable to cavity-based systems where their application requires \latin{a priori} hypotheses on the linear dispersion and 
the non-linear interactions of involved modes~\cite{Patera2010,Jiang2012}.
Such a limitation makes traditional symplectic approaches inadequate to treat a wide class of relevant experimental situations, 
including multimode features in resonant systems exploiting third order non-linear interactions.
This is, for instance, the case of important platforms for integrated quantum photonics such as silicon and silicon nitride~\cite{Kues2017,Vaidya2019}.

In this paper, we provide, a generalized strategy that extends the standard symplectic approach and permits to retrieve, 
without any assumptions or restrictions, the super-mode structure for any quadratic Hamiltonian. We consider here a generic 
below-threshold resonant system that can present linear and non-linear dispersive effects.
Our method applies to multiple scenarios.
These encompass low-dimensional systems, such as single- or double-mode squeezing in detuned devices~\cite{Fabre1990, Solimeno2002} 
or in opto-mechanical cavities~\cite{Mancini1994}, as well as highly multimode states, such as those produced via four-wave mixing 
in integrated systems on silicon photonics~\cite{Kues2017}.
Eventually, we note that the tools developed here for resonant systems can be equally used for the analysis of spatial propagation 
in single pass configurations~\cite{Lipfert2018, DavidBarral2019}.

\section{Multimode Langevin Equations}
We consider the most general time-independent quadratic Hamiltonian describing the dynamics of $N$ bosonic modes, $\hat{a}_n$, in the interaction picture:
\begin{equation}\label{ham}
H=\hbar\sum_{m,n} G_{m,n} a_m^{\dag}a_n +\frac{\hbar}{2}\sum_{m,n}\left[F_{m,n} a_m^{\dag}a_n^{\dag} + \hc{}\right].
\end{equation}
In this expression, the matrix $F$ is a complex symmetric matrix, $F=F^\transp$, while $G$ is a hermitian complex matrix, 
verifying $G=G^{\dagger}$\footnote{We are using the following notation: ${[\cdot]}^\transp$ for the transpose, ${[\cdot]}^*$ 
for the complex conjugate and ${[\cdot]}^\dagger$ for the Hermitian transpose.}.
Bosonic operators, $a_n$ and $a^\dagger_n$, satisfy the commutation relations $[ a_n, a^\dagger_m]=\delta_{n,m}$ and $[a_n, a_m]=0$.
In practical situations, the matrix $F$ is of the same kind as the one describing spontaneous parametric down-conversion 
in $\chi^{(2)}$ or in $\chi^{(3)}$ interactions, under the approximation of undepleted pumps~\cite{Jiang2012, Chembo2016}.
The very general shape of the matrix $G$ permits to take into account frequency conversion processes~\cite{Christ2011},
self- and cross-phase modulation in $\chi^{(3)}$ media~\cite{Chembo2016} and, in resonant systems,
it can also include the mode detunings from perfect resonance and dispersive effects.

For a cavity-based system, bosonic operators $\hat{a}_n$ and $\hat{a}^\dagger_n$ label the intra-cavity modes.
In the Heisenberg representation, the Hamiltonian operator $H$ permits to derive a set of coupled quantum Langevin equations 
describing the dynamics of the system observables below the oscillation threshold.
In terms of amplitude and phase quadratures, $x_n=\frac{1}{\sqrt{2}}(a^\dagger_n+a_n)$ and $y_n=\frac{\mathrm{i}}{\sqrt{2}}(a^\dagger_n-a_n)$,
Langevin equations read, in a compact matrix form:
\begin{equation}\label{lang_R}
\dv{\bm{R}(t)}{t}=
\left(-\gamma \mathbb{I} + \mathcal{M} \right)\bm{R}(t) + \sqrt{2\gamma}\, \bm{R}_{\mathrm{in}}(t).
\end{equation}
In the previous expression, $\bm{R}(t)={\left(x_1(t),\ldots x_N(t)|y_1(t),\ldots y_N(t)\right)}^{\transp}$
is a column vector of quadrature operators and $\mathbb{I}$ the identity matrix of $\mathbb{R}^{2N\times 2N}$.
As usual, $\gamma$ is the cavity damping coefficient and $\bm{R}_{\mathrm{in}}(t)$ is the quadrature vector of the input modes entering the system via the losses.
We stress that the quadratures of the cavity output fields, $\bm{R}_{\mathrm{out}}(t)$,
can be straightforwardly obtained with the input-output relations $\bm{R}_{\mathrm{in}}(t)+\bm{R}_{\mathrm{out}}(t)=\sqrt{2\gamma}\,\bm{R}(t)$~\cite{GardinerZoller}.
The mode interaction matrix, $\mathcal{M}\in \mathbb{R}^{2N\times 2N}$, explicitly depends on the matrices $F$ and $G$ that appear in the
Hamiltonian operator~\eqref{ham} via the relation:
\begin{equation}\label{Emme}
\mathcal{M}=
\left(\begin{array}{c|c}
	\Im[G-F]& \Re[G-F]
	\\ \hline 
	-\Re[G+F] & -{\Im[G-F]}^\transp
\end{array}\right),
\end{equation}
where matrices $\Re[G-F]$ and $\Re[G+F]$ are both symmetric.
We note that the system threshold is defined by the highest eigenvalue $\lambda_0$ of $\mathcal{M}$
for which $\Re[\lambda_0]=\gamma$.

As explained, finding the system super-modes corresponds to identifying the linear combinations of the original $a_n$ and $a^\dagger_n$ 
that permit to diagonalize $\mathcal{M}$, so as to uncouple the evolution equations, while preserving the symplectic structure of the problem~\cite{Valcarcel2006,Patera2010}.
However, in general $\mathcal{M}$ cannot be diagonalized by symplectic unitary transformation, a part from special cases for which the matrix $G$ 
is null~\footnote{Note that, in principle, special cases can exist for which $\mathcal{M}$ could be block-diagonalized or 
put into a canonical Jordan form via symplectic and unitary matrices.}; 
besides low-dimension systems whose equations can be solved directly~\cite{Vaidya2019}, this confines the analysis to systems presenting $\chi^ {(2)}$ 
non-linearities and mode-independent detuning~\cite{Jiang2012}.
These limitations arise from the fact that standard symplectic diagonalization methods consider a discrete number of boson operators, neglecting their 
explicit dependence on time or frequency.
Conversely, in a general situation, as we are analyzing, pertinent transformations are matrix-valued functions of frequency/time and demand an adequate 
extension of symplectic approach.

\section{Generalized symplectic approach}
As a first step, we show that, even in the most general case considered here, the transformation associated with Eqs.~\eqref{lang_R} 
and connecting the input and output modes is indeed symplectic in a more general sense.
By doing so, we can then apply to it a generalized version of Bloch-Messiah Decomposition.

Steady state solutions of Eqs.~\eqref{lang_R} can be obtained in the frequency domain by application
of the Fourier transform to the slowly-varying envelopes~\cite{GardinerZoller}:
\begin{equation}
\bm{R}(\omega)=\frac{1}{\sqrt{2\pi}}\int_{-\infty}^{+\infty}
\mathrm{e}^{-\mathrm{i}\omega t} \bm{R}(t)\,\dd{t}.
\end{equation}
The quadratures of the output modes read as:
\begin{equation}
\bm{R}_{\mathrm{out}}(\omega)=
S(\omega)\bm{R}_{\mathrm{in}}(\omega),
\label{R_sol}
\end{equation}
where $S(\omega)$ is the transfer function of the linear system~\eqref{lang_R}:
\begin{equation}
S(\omega)=\frac{2\gamma}
{\left(\mathrm{i}\omega+\gamma\right)\mathbb{I}-\mathcal{M}}-\mathbb{I}.
\end{equation}
This is a complex matrix-valued function, verifying $S(-\omega)=S^*(\omega)$, which assures the realty of S in time domain.
In matrix form, the commutators of input mode quadratures can be written as
$\left[\bm{R}_{\mathrm{in}}(\omega),\bm{R}_{\mathrm{in}}^{\transp}(\omega')\right]=\Omega\,\delta(\omega+\omega')$
where $\Omega=
\begin{pmatrix}
	0 & I \\
	-I & 0
\end{pmatrix},
$
is the N-mode \emph{symplectic form} and $I$ the identity matrix of $\mathbb{R}^{N\times N}$~\cite{Adesso2014}.
In order to guarantee that the commutators are preserved for $\bm{R}_{\mathrm{out}}(\omega)$, the transformation $S(\omega)$ must verify
\begin{equation}
\forall\; \omega\in\mathbb{R}:\quad
S(\omega)\,\Omega\, S^{\transp}(-\omega)=\Omega.
\label{w-SOS}
\end{equation}
In the case we are dealing with, this condition is easily verified (see Appendix A) by noticing that the matrix $\mathcal{M}$ of Eq.~\eqref{Emme} is a \emph{Hamiltonian matrix},
\latin{i.e.\@} it verifies  the relation ${(\Omega\,\mathcal{M})}^{\transp}=\Omega\,\mathcal{M}$.
%
Expression~\eqref{w-SOS} extends the standard symplecticity condition as known in the literature~\cite{Adesso2014}.
More precisely, it defines a set of transformations that depend on a real continuous parameter -- the frequency $\omega$ --
and such that every matrix obtained from $S(\omega)$ with $\omega$ assigned belongs to the \emph{conjugate symplectic group} $Sp^*(2N,\mathbb{C})$~\cite{Mackey2003}:
\begin{equation}
\mathbb{S}_{\omega}=\left\{ S(\omega) \in \mathscr{C}_\omega\; \big|\; \forall \,\omega\in\mathbb{R}, S(\omega)\in Sp^*(2N,\mathbb{C})\right\},
\end{equation}
where $\mathscr{C}_\omega$ is the set of matrix-valued functions in $\mathbb{C}^{2N\times2N}$ that are smooth with respect to $\omega$.
For the sake of simplicity, we will refer to transformation belonging to $\mathbb{S}_{\omega}$ as \emph{$\omega-$symplectic}.

In a general way, $\omega$-symplectic transformations admit a decomposition that is a smooth function of the real parameter, 
as expected to describe the mode continuous evolution in time/frequency.
In other words, for any element of $\mathbb{S}_{\omega}$ it exists an \emph{analytical Bloch-Messiah Decomposition} (ABMD, see Appendix B):
\begin{equation}
S(\omega)=U(\omega) D(\omega) V^\dagger(\omega),
\label{w-BM}
\end{equation}
where $U(\omega)$, $D(\omega)$, and $V(\omega)$ are smooth matrix-valued functions such that,
for any assigned value of $\omega$, $U(\omega),V(\omega)\in Sp^*(2N,\mathbb{C})\cap \mathcal{U}(2N)$, with $\mathcal{U}(2N)$ the unitary group.
The matrix $D(\omega)= \diag\{d_1(\omega),\ldots,d_N(\omega)|\,d_1^{-1}(\omega),\ldots,d_N^{-1}(\omega)\}$ with $d_m(\omega)\ge1$ for $m=\{1,\ldots,N\}$,
for all $\omega\in\mathbb{R}$.
We note that these matrix-valued functions can be chosen, after conjugating Eq.~\eqref{w-BM}, so to verify the same property as $S^*(\omega)=S(-\omega)$.

Expression~\eqref{w-BM} shows that a BMD for $S(\omega)$, in the case of a generic quadratic Hamiltonian, exists and depends on a continuous parameter.
From it, the quadrature of super-modes of system~\eqref{lang_R} can be obtained as $\bm{R}'_{\mathrm{out}}(\omega)=U^{\dagger}(\omega)\bm{R}_{\mathrm{out}}(\omega)$,
where we have assumed input vacuum state.
We remark that the shape of the super-modes themselves depends on the continuous parameter:
this result shows that in practical situations, the optimal detection modes change with the analysis frequency, $\omega$.

To conclude, we note that Eqs~\eqref{w-SOS} and~\eqref{w-BM} have counterparts in the time domain.
The matrix-valued Green function $S(t)$ of~\eqref{lang_R},
corresponding to the inverse Fourier transform of $S(\omega)$, is symplectic in the sense that $\forall\; t,t'\in\mathbb{R}$:
\begin{equation}
\int_{-\infty}^{+\infty} S(t-\tau)\,\Omega\,S^\transp(t'-\tau) \, \dd{\tau}
	= \Omega \,\delta (t-t'),
\label{t-SOS}
\end{equation}
and its ABMD reads:
\begin{equation}
S(t)
	= \int_{-\infty}^{+\infty} U(\tau)\, D(t-\tau+\tau')\, V(\tau') \, \dd{\tau}\dd{\tau'},
\label{t-BM}
\end{equation}
where $U(t)$ and $V(t)$ real matrix-valued Green functions.
They are symplectic in the sense of~\eqref{t-SOS}
and orthogonal in the sense $(U\star U^{\transp})(t)=\mathbb{I}\,\delta(t)$ and $(V\star V^{\transp})(t)=\mathbb{I}\,\delta(t)$,
with $\star$ the cross-correlation product.
The Green function $D(t)$ is the diagonal matrix-valued obtained as the inverse
Fourier transform of $D(\omega)$.
It is real and even since $D(\omega)$ is real and even.

\section{Spectrum of quantum noise}
We now characterize the quantum statistical properties of the output steady states $\bm{R}_{\mathrm{out}}$
and of their super-modes $\bm{R}'_{\mathrm{out}}$.
To this purpose, we consider a generic linear combination $Z_{\bm{\theta}}$ of $\bm{R}_{\mathrm{out}}$ specified by
the normalized line-vector $\bm{Q}(\bm{\theta})$ consisting of real coefficients:
\begin{equation}
Z_{\bm{\theta}}
	= \bm{Q}(\bm{\theta})\;\bm{R}_{\mathrm{out}},
\label{Ztheta}
\end{equation}
where $\bm{\theta}$ are the $2N-1$ angles parametrizing $\mathbf{Q}(\bm{\theta})$.

The spectrum of quantum noise can be expressed by means of the Wiener-Khinchin theorem in terms of the self-correlation of $Z_{\bm{\theta}}$ as:
\begin{equation}
\Sigma_{\bm{\theta}}(\omega)
	= \int_{-\infty}^{+\infty} \mathrm{e}^{-\mathrm{i}\omega\tau} \,
		\left\langle Z_{\bm{\theta}}(t+\tau)Z_{\bm{\theta}}(t)\right\rangle\, \dd{\tau}.
\label{sqz_theta}
\end{equation}
By making use of expression~\eqref{Ztheta} in the frequency domain, Eq.~\eqref{sqz_theta} can be written as:
\begin{equation}
\Sigma_{\bm{\theta}}(\omega)
	= \bm{Q}(\bm{\theta})\,\sigma_{\mathrm{out}}(\omega)\,{\bm{Q}(\bm{\theta})}^{\transp}
\label{sqz_spectrum}
\end{equation}
where
\begin{equation}
\sigma_{\mathrm{out}}(\omega)
	= \frac{1}{2\sqrt{2\pi}}S(\omega)S^{\transp}(-\omega)
\label{sig-SS}
\end{equation}
is the Fourier transform of the covariance matrix of the output state $\sigma_{\mathrm{out}}(\tau)
	= \frac{1}{2}\langle \bm{R}_{\mathrm{out}}(0)\bm{R}_{\mathrm{out}}^{\transp}(\tau)
		+{(\bm{R}_{\mathrm{out}}(0)\bm{R}^{\transp}_{\mathrm{out}}(\tau))}^{\transp}\rangle$,
that depends only on time differences, $\tau$, as we are considering a stationary regime~\cite{Kolobov2011}.
In Eq.~\eqref{sig-SS}, we used~\eqref{R_sol} and the fact that
for vacuum input state $\sigma_{\mathrm{in}}(\tau)=\frac{\mathbb{I}}{2}\delta(\tau)$.

Equation~\eqref{sig-SS} can be re-written by making use of the ABMD in Eq.~\eqref{w-BM}.
We obtain:
\begin{equation}
\sigma_{\mathrm{out}}(\omega)
	= \frac{1}{2\sqrt{2\pi}}U(\omega)D^2(\omega)U^{\dag}(\omega).
\label{sigBMD}
\end{equation}
By replacing~\eqref{sigBMD} into~\eqref{sqz_spectrum} it is clear that, in general, optimal squeezing (resp.\@ anti-squeezing) cannot be
reached by any linear combination $\bm{Q}(\bm{\theta})$ apart from those cases where $U(\omega)$ is real.
In this case optimality could be reached only at a given value of $\omega$, by choosing $\bm{Q}(\bm{\theta})$
equal to one column of $U(\omega)$, as we will show in the next section.
In experiments, the super-modes properties, and in turn their squeezing features, can be obtained by replacing
$\bm{Q}(\bm{\theta})$ by a complex line vector-valued function $\bm{Q}(\bm{\theta}(\omega))=U^\dagger(\omega)$.
With this choice,
\begin{equation}\label{Pene}
\Sigma_{\bm{\theta}}(\omega) = \frac{1}{2\sqrt{2\pi}}D^{2}(\omega).
\end{equation}
Based on this expression, the elements of the diagonal matrix $D^{2}(\omega)$ give directly the variances of super-modes quadratures and they can be interpreted as
their anti-squeezing $\{d_1(\omega),\ldots,d_N(\omega)\}$ and squeezing levels $\{d_1^{-1}(\omega),\ldots,d_N^{-1}(\omega)\}$.
We note that assigning $\mathbf{Q}(\bm{\theta}(\omega))$ corresponds to designate a particular shape of the local oscillator
(LO) of a homodyne detection scheme.
As a consequence, in order to retrieve the optimal information on super-modes, the LO itself must depend on $\omega$ and be chosen according to the analysing frequency.
The shaping of the LO could be implemented, for example, by a passive interferometer with memory effect.

\section{Single mode squeezing in detuned optical cavity}
The case of a single mode squeezed state generated in a detuned optical parametric oscillator (OPO) is already illustrative of the relevance of a 
continuous-parameter symplectic approach.
In this case the vector of field quadratures is $\bm{R}={(x,y)}^\transp$ and the matrix $\mathcal{M}$ associated to this system is:
\begin{equation}
\mathcal{M}=
\begin{pmatrix}
	g & \Delta \\
	-\Delta & -g
\end{pmatrix},
\end{equation}
where $g$ accounts for the parametric gain and $\Delta$ is the detuning from cavity resonance of the squeezed mode.

The system has two singular values $d_1(\omega)$ and $d_1^{-1}(\omega)$ and, associated to these, two super-modes.
As the super-mode quadratures are found to have real coefficients, we can write them as:
$R'_{\mathrm{out},i}=\cos[\theta_i(\omega)]\,x_\mathrm{out}+\sin[\theta_i(\omega)]\,y_\mathrm{out}$ with $i=1,2$.
The quadrature angles are frequency dependent and verify $\theta_2(\omega)=\theta_1(\omega)+\pi/2$.

\begin{figure}[t!]
\centering
\includegraphics[width=0.45\linewidth]{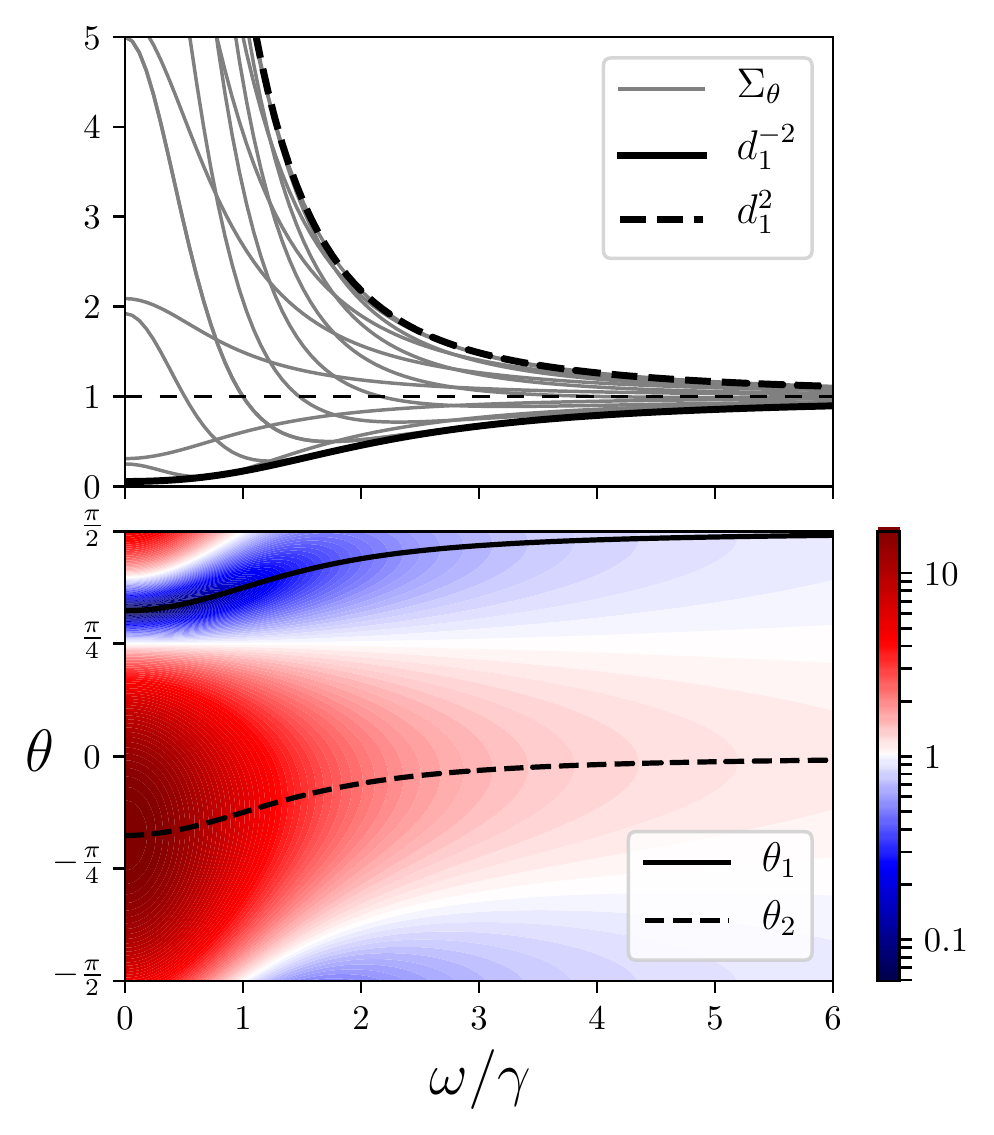}
\caption{Top: Frequency-dependent singular values $d_1^2(\omega)$ and $d_1^{-2}(\omega)$.
They quantify the degree of anti-squeezing (dashed black) and squeezing (solid black) respectively.
Solid grey curves represent the normalized to SQL
spectrum of quantum noise $\Sigma_{\theta}$ of quadratures $Z_{\theta}$ for several values of $\theta$.
Bottom: color density plot represents $\Sigma_{\theta}$ with respect to analysis frequency $\omega$ and quadrature angle $\theta$.
Dashed and solid black lines represent the frequency dependent angles $\theta_i(\omega)$ associated to super-modes $R'_{\mathrm{out},i}$.
They represent the set of points in the space $(\omega,\theta)$
for which $\Sigma_{\theta}$ is minimum (squeezing) or maximum (anti-squeezing).
}\label{Fig1}
\end{figure}
In figure~\ref{Fig1}-top, we trace $d_1^2(\omega)$ (solid) and $d_1^{-2}(\omega)$ (dashed), as functions of the analysis frequency $\omega$ and 
we compare them to the Standard Quantum Limit (SQL).
The figure also shows (in gray) normalized-to-SQL spectra $\Sigma_{\theta}$ of field quadratures,
$Z_{\theta}=\cos \theta\,x_\mathrm{out}+\sin\theta\,y_\mathrm{out}$, calculated for several values of the angle $\theta$, with $\theta$ frequency independent.
These quadratures are obtained by imposing in Eq.~\eqref{Ztheta} a real and constant $\bm{Q}(\bm{\theta})$.
Regardless the choice of $\theta$, the curves $\Sigma_{\theta}$ exhibit a (local or asymptotic) minimum but do not reach the optimal squeezing for all values of $\omega$.
Conversely, the function $d_1^{-2}(\omega)$ corresponds the envelope of $\Sigma_{\theta}$ minima, thus confirming that the optimal squeezing spectrum
is the one computed for the super-modes.
A similar observation holds for the anti-squeezing, $d_1^{2}(\omega)$.

Figure~\ref{Fig1}-bottom shows the angles $\theta_1(\omega)$ and $\theta_2(\omega)$ that give the super-mode coefficients.
The color code indicates quadrature noise levels normalized to SQL as functions of $\omega$ and of the quadrature angle $\theta$.
As expected, when $\omega$ changes, the frequency dependent angles, $\theta_1(\omega)$ and $\theta_2(\omega)$ associated to super-modes correctly
gives the superpositions of $x_{\mathrm{out}}$ and $y_{\mathrm{out}}$ that lead to optimal anti-squeezing and squeezing levels.
We note that the dependence of the optimal quadrature angle with respect to analysis frequency is in agreement with the result obtained
by directly solving the one-dimension Langevin equations either in detuned OPO~\cite{Fabre1990} or optomechanical
cavities~\cite{Fabre1990,Mancini1994}.

\section{Four-mode system}
\begin{figure}[t!]
\centering
\includegraphics[width=0.45\linewidth]{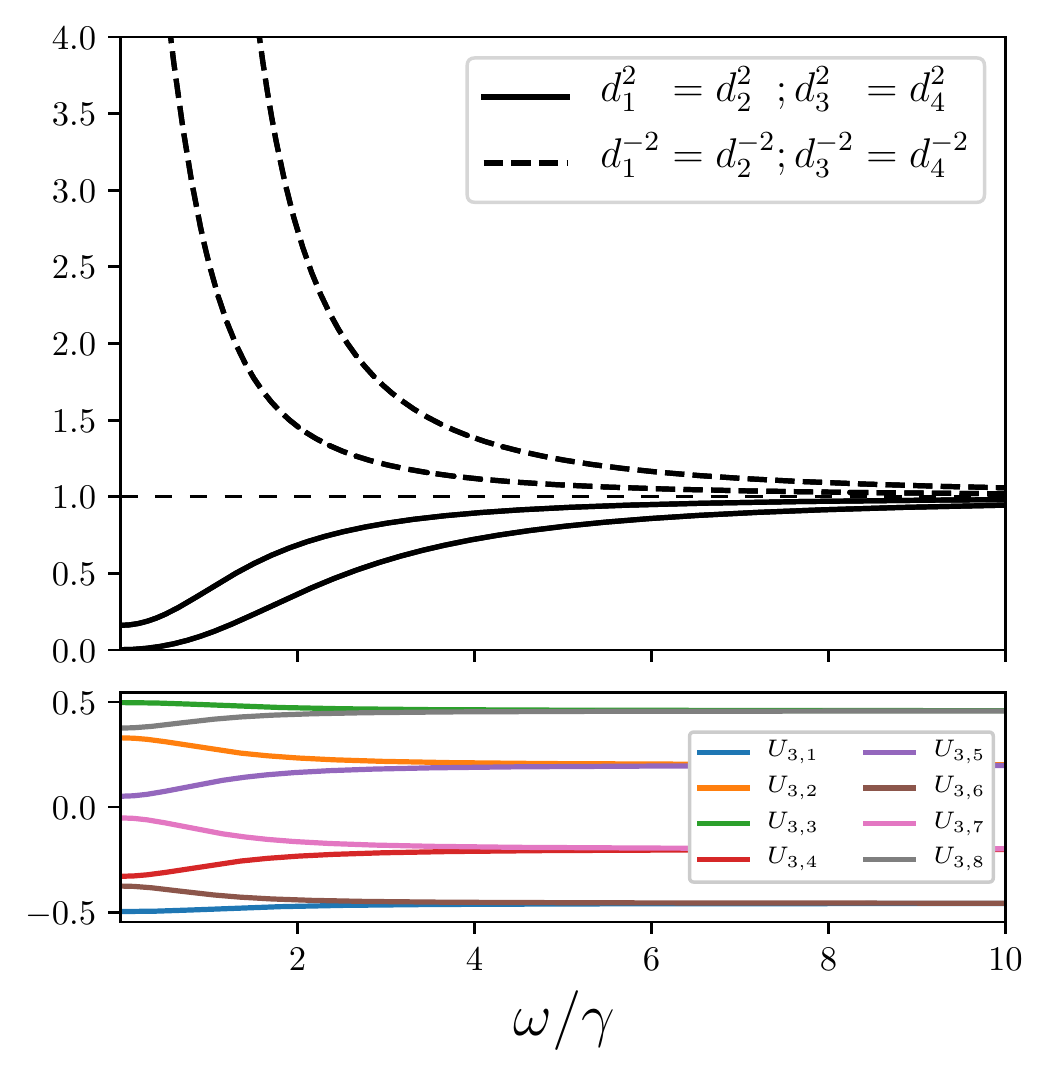}\vspace{-0.4cm}
\caption{Top: frequency dependent singular values $d_i(\omega)$ and $d_i^{-2}(\omega)$,
for $i=1,\ldots,4$, quantifying the degree of anti-squeezing (dashed) and squeezing (solid) of corresponding super-mode.
Level of noise equal to one corresponds to SQL\@.
Bottom: frequency dependent (real) coefficients $U_{3,j}(\omega)$ of one ($i=3$) of the eight super-modes obtained from ABMD.}\label{Fig2}
\end{figure}
To conclude we discuss a case that is complex enough to demonstrate the efficacy of the generalized symplectic approach and the ABMD\@.
We chose a multimode system with $N=4$ in the case of both $F$ and $G$ non null.
The structure of these two matrices is chosen as
\begin{equation}
F=\left(\begin{array}{c|c}
   \tilde{F} & 2\tilde{F}
   \\
   \hline
   2\tilde{F} & \tilde{F}
  \end{array}\right),
\quad
G=\left(\begin{array}{c|c}
   2\tilde{G} & \tilde{G}
   \\
   \hline
   \tilde{G} & 2\tilde{G}
  \end{array}\right),
\end{equation}
with $\tilde{F}=\begin{pmatrix} 0 & a \\ a & 0 \end{pmatrix}$ and $\tilde{G}=\begin{pmatrix} b & 0 \\ 0 & b \end{pmatrix}$.
This scenario is, for instance, the one of a $\chi^{(3)}$ process driven by two strong pumps that gives origin to both parametric and frequency conversions,
including self- and cross- phase modulation of signal and idler waves.

For this system the ABMD  gives 8 singular values and 8 super-modes that are smooth with respect to $\omega$.
In figure~\ref{Fig2}-top we trace the frequency-dependent singular values that, for this specific case, are two by two degenerate.
The solid lines represent the square of $d_i(\omega)$ (resp. $d_i^{-1}(\omega)$) for $i=1,\ldots,4$
and they are compared to SQL\@.
They correctly provide the minimum (resp.\@ maximum) degree of squeezing (resp.\@ anti-squeezing) produced by the system at a given
value of the analysis frequency $\omega$.
In figure~\ref{Fig2}-bottom we represent the 8 frequency-dependent coefficients of one of the super-modes ($i=3$).

\section{Conclusions}
We developed a generalized symplectic approach that allows to tackle the general problem of the identification of the super-mode structure 
for any system evolving under a generic quadratic Hamiltonian in a cavity-based configuration.
The presented strategy allows to cover the analysis of many optical systems that are relevant for quantum technologies but that cannot 
be easily analyzed by standard symplectic diagonalizations.
In this framework we introduced the analytical Bloch-Messiah decomposition, extending traditional methods to symplectic transformations 
that depend on a continuous parameter such as the frequency $\omega$.
As a result of the decomposition, super-modes  and their associated singular values are, in the most general case, dependent of the continuous parameter.
Our approach will allow to treat easily problems with very large number of degrees of freedom, hence enabling a better harvesting and control of their quantum properties.
This feature is of crucial importance for application in the domain of quantum technologies with a major impact in the development of bulk and integrated quantum optics.

\acknowledgments
G.P. acknowledges useful discussions with L. Dieci and L. Lopez.

\appendix
\section{Proof that $S(\omega)$ is $\omega$-symplectic}
For any $\omega$ we have
\begin{equation}
S(\omega)=\frac{\left(\gamma-\mathrm{i}\omega\right)\mathbb{I}+\mathcal{M}}{\left(\gamma+\mathrm{i}\omega\right)\mathbb{I}-\mathcal{M}}.
\label{Som}
\end{equation}
We note that the order in which is taken the numerator and denominator does not matter since they commute.
We use the property of $\mathcal{M}$ of being a Hamiltonian matrix ${\left(\Omega\mathcal{M}\right)}^{\transp}=\Omega\mathcal{M}$.
Then we evaluate
\begin{align}
S(\omega)\;\Omega\; {S(-\omega)}^{\transp}&=
\frac{\left(\gamma-\mathrm{i}\omega\right)\mathbb{I}+\mathcal{M}}{\left(\gamma+\mathrm{i}\omega\right)\mathbb{I}-\mathcal{M}}\;
\Omega\;
\frac{\left(\gamma+\mathrm{i}\omega\right)\mathbb{I}+\mathcal{M}^{\transp}}{\left(\gamma-\mathrm{i}\omega\right)\mathbb{I}-\mathcal{M}^{\transp}}
\end{align}
Then by replacing $\mathbb{I}=\Omega^{-1} \mathbb{I} \Omega$ and $\mathcal{M}^{\transp}=-\Omega^{-1}\mathcal{M}\Omega$ in the second term on their right-hand side, we obtain
\begin{align}
S(\omega)\;\Omega\; {S(-\omega)}^{\transp}&=
\frac{\left(\gamma-\mathrm{i}\omega\right)I+\mathcal{M}}{\left(\gamma+\mathrm{i}\omega\right)I-\mathcal{M}}\;
\frac{\left(\gamma+\mathrm{i}\omega\right)I-\mathcal{M}}{\left(\gamma-\mathrm{i}\omega\right)I+\mathcal{M}}
\Omega=\Omega
\end{align}
which proves that $S(\omega)$ is a conjugate-symplectic matrix for every $\omega$.

\section{Proof of the existence of analytical Bloch-Messiah decomposition}
In~\cite{Bunse-Gerstner1991,Wright1992,Dieci1999} constructive methods for finding the analytic singular value decomposition 
of a matrix smoothly depending on a real parameter are given, while in~\cite{Dieci2006} the analytic singular value decomposition on the real symplectic group
has been considered.
In the case with no degenerate singular values, at each $\omega$ the Bloch-Messiah decomposition is unique up to the order of the singular values and vectors, and up to a phase for each singular vector.
Those vectors form a conjugate-symplectic base.
Let $x(\omega)$ and $y(\omega)$ be two normed eigenvectors given by the application of smooth decomposition without taking into account symplecticity.
By quasi-unicity of the singular value decomposition, up to a phase they are part of a conjugate-symplectic base, $\abs{x^\dagger \Omega y}$ can only take one of the values $0$ or $1$.
As $\omega \mapsto x(\omega)$ and $\omega \mapsto y(\omega)$ are continuous, $f: \omega \mapsto \abs{x^\dagger(\omega) \Omega y(\omega)}$ is also continuous.
Assuming a connected domain for $\omega$ implies that $f$ is constant.
Phase can be continuously corrected when needed by replacing $y$ with $\dfrac{y}{x^\dagger \Omega y}$.
This being true for all possible pairs of $x$ and $y$, we conclude that if for a given $\omega$ the passage matrix is conjugate-symplectic, it keeps this property for all $\omega$.
We call it an analytic Bloch-Messiah decomposition.
In this section we show how to constructively express the decomposition for a given $S(\omega)\in\mathbb{S}_{\omega}$ of the following form:
\begin{equation}
S(\omega)=U(\omega)D(\omega)V^{\dagger}(\omega)
\label{UDV}
\end{equation}
where $U$ and $V$ are $2N\times 2N$ unitary and conjugate-symplectic matrix-valued functions
and $D$ is a $2N\times 2N$ diagonal matrix-valued function
\begin{equation}
D(\omega)=\begin{pmatrix}
          D_1(\omega) & 0
          \\
          0 & D_2(\omega)
          \end{pmatrix}
\end{equation}
were $D_1(\omega)=\text{diag}\{d_1(\omega),\ldots,d_N(\omega)\}$ and $D_2(\omega)=\text{diag}\{d_1^{-1}(\omega),\ldots,d_N^{-1}(\omega)\}$
with $d_k(\omega)>0$ and $1 \leq d_k(0)\leq d_{k+1}(0)$ (we note that for $\omega>0$ the order of the singular values can change with respect to the
initial one for an analytical decomposition~\cite{Bunse-Gerstner1991}).
It is easy to prove that as elements of the intersection between
the conjugate-complex symplectic and the unitary groups the matrices $U$ and $V$ have the following block-form
\begin{equation}
U(\omega)=\begin{pmatrix}
          U_1(\omega) & U_2(\omega)
          \\
          -U_2(\omega) & U_1(\omega)
          \end{pmatrix}
\quad\text{and}\quad
V(\omega)=\begin{pmatrix}
          V_1(\omega) & V_2(\omega)
          \\
          -V_2(\omega) & V_1(\omega)
          \end{pmatrix}
\end{equation}
By differentiating~\eqref{UDV} with respect to $\omega$ (we designate the symbol $'$ for derivation with respect to $\omega$ and 
temporary drop the dependence on $\omega$ for space-saving):
\begin{equation}
S'=U' D V^{\dagger}+U D'V^{\dagger}+U D V'^{\dagger}.
\label{UDV2}
\end{equation}
After multiplying~\eqref{UDV2} by $U^\dagger$ from the right and by $V$ from the left
\begin{equation}
D'=U^\dagger S' V - U^\dagger U' D -  D V'^\dagger V
\label{UDV3}
\end{equation}
Now we define $H=U^\dagger U'$ and $K=V^\dagger V'$ and, then, we multiply from the left these definitions by $U$ and $V$ and we get
\begin{align}
U' &= U H, \label{Uprime}
\\
V' &= V K. \label{Vprime}
\end{align}
Since $U$ and $V$ are unitary, then
\begin{align}
U^\dagger U &= I \label{UdagU},
\\
V^\dagger V &= I \label{VdagV}.
\end{align}
After differentiating~\eqref{UdagU} and~\eqref{VdagV}, we find that $H^\dagger=-H$ and $K^\dagger=-K$ are anti-Hermitian.
Moreover, by construction, $H$ and $K$ have the block-structure
\begin{equation}
H=
\begin{pmatrix}
H_1 & H_2
\\
-H_2 & H_1
\end{pmatrix}
\quad
\text{and}
\quad
K=
\begin{pmatrix}
K_1 & K_2
\\
-K_2 & K_1
\end{pmatrix}.
\end{equation}
These properties guarantee that the matrices $H$ and $K$ are Hamiltonian matrices in the sense $\Omega H = {(\Omega H)}^\dagger$ and $\Omega K = {(\Omega K)}^\dagger$ 
and, as a consequence, that the solutions $U$ and $V$ of~\eqref{Uprime} and~\eqref{Vprime} are conjugate-symplectic and unitary matrices.

On the other side, we define $Q=U^\dagger S' V$, then re-write~\eqref{UDV3}
as
\begin{equation}
D'=Q - H D + D K
\label{Dprime}
\end{equation}
Eqs.~\eqref{Uprime},~\eqref{Vprime} and~\eqref{Dprime} define a system of differential equations for the elements of $U(\omega)$, $V(\omega)$ and $D(\omega)$
that we endow with the initial conditions $U(0)$, $V(0)$ and $D(0)$ obtained from the Bloch-Messiah decomposition at $\omega=0$
\begin{equation}
S(0)=U(0)D(0)V^\dagger(0).
\end{equation}
We now re-write Eqs.\eqref{Dprime} in the block structure
\begin{align}
\begin{pmatrix}
D_1' & 0
\\
0 & D_2'
\end{pmatrix}
&=
\begin{pmatrix}
Q_1 & Q_2
\\
Q_3 & Q_4
\end{pmatrix}
-
\begin{pmatrix}
H_1 & H_2
\\
-H_2 & H_1
\end{pmatrix}
\begin{pmatrix}
D_1 & 0
\\
0 & D_2
\end{pmatrix}
+
\begin{pmatrix}
D_1 & 0
\\
0 & D_2
\end{pmatrix}
\begin{pmatrix}
K_1 & K_2
\\
-K_2 & K_1
\end{pmatrix}
\label{Dprime_II}
\end{align}
This expression give rise to two sets of differential equations for the singular values
\begin{align}
D_1' &= Q_1 - H_1 D_1 + D_1 K_1,
\label{Dprime_A}
\\
D_2' &=   Q_4 - H_1 D_2 + D_2 K_1
\label{Dprime_B}
\end{align}
and two sets of algebraic equations
\begin{align}
H_2 D_2 - D_1 K_2 &= Q_2 ,
\label{Dprime_C}
\\
H_2 D_1 - D_2 K_2 &= - Q_3.
\label{Dprime_D}
\end{align}
First we solve Equations~\eqref{Dprime_C} and~\eqref{Dprime_D} in $H_2$ and $K_2$.
For $i,j=1,\ldots,N$
\begin{align}
{(H_2)}_{ij} d_j^{-1} - {(K_2)}_{ij} d_i  = {(Q_2)}_{ij},
\\
{(H_2)}_{ij} d_j - {(K_2)}_{ij} d_i^{-1}  = -{(Q_3)}_{ij}.
\end{align}
For $i=j$ and $d_i\neq 1$ the solutions are
\begin{align}
{(H_2)}_{ii}&=\frac{{(Q_3)}_{ii}d_i^3+{(Q_2)}_{ij}d_i}{1-d_i^4},
\label{H2ii}
\\
{(K_2)}_{ii}&=\frac{{(Q_2)}_{ii}d_i^3+{(Q_3)}_{ii}d_i}{1-d_i^4}.
\label{K2ii}
\end{align}
In the case where $d_i=1$ it must be ${(Q_3)}_{ii}=-{(Q_2)}_{ii}$.
As a consequence the system is underdetermined and we have the freedom to choose ${(K_2)}_{ii}=0$.
Hence ${(H_2)}_{ii}={(Q_2)}_{ii}$.
For $i\neq j$, $d_i\neq1$ and $d_j\neq 1$ (remember $1 \leq d_i$) the solutions are
\begin{align}
{(H_2)}_{ij}&=\frac{{(Q_3)}_{ij}d_i^2 d_j+{(Q_2)}_{ij}d_j}{1-d_i^2 d_j^2},
\label{H2ij}
\\
{(K_2)}_{ij}&=\frac{{(Q_2)}_{ij}d_i^2 d_j+{(Q_3)}_{ij}d_j}{1-d_i^2 d_j^2}.
\label{K2ij}
\end{align}
Otherwise if $d_i=d_j=1$ it must be ${(Q_3)}_{ij}=-{(Q_2)}_{ij}$ and the underdetermined system allows us to choose
${(K_2)}_{ij}=0$ and ${(H_2)}_{ij}={(Q_2)}_{ij}$.

From Eqs.~\eqref{Dprime_A} and its Hermitian conjugate we consider first the case $i=j$, for $i=1,\ldots,N$:
\begin{align}
d_i'&={(Q_1)}_{ii}-{(H_1-K_1)}_{ii}d_i,
\label{dprime1}
\\
d_i'&={(Q_1)}_{ii}^*+{(H_1-K_1)}_{ii}d_i.
\label{dprime1 bis}
\end{align}
By summing Eqs.~\eqref{dprime1} and~\eqref{dprime1 bis} we get a set of differential equation for the singular values
\begin{equation}
d_i'=\frac{{(Q_1)}_{ii}+{(Q_1)}_{ii}^*}{2}.
\label{dprime}
\end{equation}
By subtracting Eqs.~\eqref{dprime1} and~\eqref{dprime1 bis} we get an algebraic equation that allows to obtain the diagonal
elements of $H_1$ and $K_1$:
\begin{equation}
{(H_1-K_1)}_{ii}=\frac{{(Q_1)}_{ii}-{(Q_1)}_{ii}^*}{2 d_i}.
\label{H1-K1ii}
\end{equation}
We can choose, then, ${(K_1)}_{ii}=0$ and determine ${(H_1)}_{ii}$. Notice that this result is consistent with the fact that $H$ and $K$ are skew-Hermitian
so that their diagonal must be purely imaginary.
Now we consider Eqs.~\eqref{Dprime_A} and its Hermitian conjugate for $i\neq j$, for $i,j=1,\ldots,N$.
In this case, after using the fact that $H$ and $K$ are skew-Hermitian, we get
\begin{align}
{(Q_1)}_{ij}-{(H_1)}_{ij}d_j+{(K_1)}_{ij}d_i &=0,
\label{Dprime_Aij}
\\
{(Q_1)}_{ji}^*+{(H_1)}_{ij}d_i-{(K_1)}_{ij}{d_j} &=0.
\label{Dprime_Bij}
\end{align}
We can solve Eqs.~\eqref{Dprime_Aij} and~\eqref{Dprime_Bij} with respect to ${(H_1)}_{ij}$ and ${(K_1)}_{ij}$.
This system of algebraic equation is solvable if $d_i\neq d_j$, which means that the spectrum of singular values is not degenerate.
In this case we obtain
\begin{align}
{(H_1)}_{ij}&=\frac{{(Q_1)}_{ij}d_j-{(Q_1)}_{ij}^*d_i}{d_j^2-d_i^2},
\label{H1ij}
\\
{(K_1)}_{ij}&=\frac{{(Q_1)}_{ij}d_i-{(Q_1)}_{ij}^*d_j}{d_j^2-d_i^2}.
\label{K1ij}
\end{align}
The case where the path of two or more singular values collide thus giving rise to degeneracies can also be treated by adapting to our case the
strategy developed in~\cite{Wright1992,Dieci1999} for the case of the analytic singular value decomposition. 

We notice also that in the case of transformations like \eqref{Som}, some of the degeneracies in the spectrum
of $S(\omega)$ can derive from degeneracies in the spectrum of the eigenvalues of $\mathcal{M}$.
In this case if a degeneracy is present at $\omega=0$ it will persist at any other $\omega\neq0$.

Finally, the algorithm that allows to find the analytical Bloch-Messiah decomposition of $S(\omega)$ is the following.
We start at $\omega_0=0$ and we find the standard BMD $S(0)=U(0)D(0)V^\dagger(0)$.
From $U(0)$, $V(0)$ and $D(0)$ we evaluate $Q(0)=U^\dagger(0)s'(0)V(0)$ as well as $H_1(0)$ and $K_1(0)$ from eqs~\eqref{H1-K1ii},~\eqref{H1ij} and~\eqref{K1ij} and $H_2(0)$ and $K_2(0)$ from the solutions of the system~\eqref{Dprime_C} and~\eqref{Dprime_D}.
Then we can find, from Euler approximation of Eq.~\eqref{dprime} the matrix $D(\omega_1)$.
On the other side, for solving Eqs.~\eqref{Uprime} and~\eqref{Vprime} we use the Magnus perturbative approach that has the advantage of preserving the symplectic structure at any order of approximation.
The solutions $U(\omega_1)$ and $V(\omega_1)$, with $\omega_1=\omega_0+d\omega$ (with $d\omega\ll1$), are thus
evaluated at the first Magnus order as:
\begin{equation}
U(\omega_1)\approx U(\omega_0)\,\exp\Big(H(\omega_0)\dd{\omega}\Big).
\end{equation}
These results are used for obtaining
the values of $H(\omega_1)$ and $K(\omega_1)$, then the procedure can be iterated for $\omega_m=\omega_0+m\, d\omega$ with $m>1$.
\bibliography{biblioHAL}

\end{document}